\documentstyle[namedreferences,psfig]{kluwer}

\let\oldverbatim\verbatim
\renewcommand{\verbatim}{\expandafter\small\oldverbatim}
\def\la{\mathrel{\hbox{\rlap{\hbox{\lower4pt\hbox{$\sim$}}}\hbox{$<$}}}}
\def\ga{\mathrel{\hbox{\rlap{\hbox{\lower4pt\hbox{$\sim$}}}\hbox{$>$}}}}
\def\fm{\hbox{$.\!\!^{m}$}}
\def\micron{\hbox{$\mu$m}}
\def\farcm{\hbox{$.\mkern-4mu^\prime$}}
\def\deg{{^\circ}}
\newcommand{\kms}{{\,km\ sec^{-1}\,}}

\runningtitle{THE PERFORMANCE OF MEFOS}
\runningauthor{P. FELENBOK ET AL.}

\begin{opening}

\title{The Performance of MEFOS, the ESO Multi-Object Fibre Spectrograph}

\author{P. \surname{Felenbok}}
\author{J. \surname{Gu\'erin}}
\author{A. \surname{Fernandez}}
\author{V. \surname{Cayatte}}
\author{C. \surname{Balkowski}}
\author{R.C. \surname{Kraan-Korteweg}}
\institute{Observatoire de Paris, DAEC, Unit\'e associ\'ee au CNRS, D0173, et 
\`a l'Universit\'e Paris 7, \\ 92195 Meudon Cedex, France}

\date{}

\end{opening}

\begin{document}

\begin{abstract}

We are describing a new multi-fibre positioner, MEFOS, that was 
in general use at the La Silla Observatory, and implemented at the prime
focus of the ESO 3.6 m telescope. It is an arm positioner using 29 
arms in a one degree field. Each arm is equipped with an individual 
viewing system for accurate setting and   carries two spectroscopic 
fibres, one for the
astronomical object and the other one for the sky recording needed for sky
subtraction. The spectral fibres  intercept 2.5 arcsec on the sky
and run from the prime focus to the Cassegrain, where the B\&C
spectrograph is located. After describing the observational procedure, we
 present the first scientific results.

\end{abstract}

\keywords{Spectrographs, Observations, Large-scale structure of the Universe}

\section{Introduction}

With the increasing demand of the astronomical community for
statistical work, all the major observatories equipped their telescopes
with spectrographs having multiple object capabilities. We can distinguish
between small field instruments, less than 10 arcmin, and large field ones,
more than one degree instruments, on which we will focus here. The large field ones use 
essentially fibre fed spectrographs. This is true for small
telescopes (like  the UK Schmidt) as well as for the 4m class ones. The first
generation instruments used  focal plates with manually inserted
optical fibres in plugboards with drilled holes made at the object
coordinates. This system is still foreseen for dedicated instruments, as
the SLOAN DSS project(Limmongkol {\it et al.} 1992), but for general use facilities, automatic fibre
handling machines are preferred. Two main devices are in operation, robot and
arm positioners. The robot positioner plugs the different fibres
in sequence whereas the arm positioner moves  all the fibres simultaneously.
The robot is able to take care of a large number of fibres, but the
configuration time is long and the positioning is done in a blind manner although the robot is equipped with a vision system. This does not guarantee the dropping process.
Arms are fast, but more than 30 are hardly feasible. Moreover, they are 
expensive, except if built in-house.

When ESO decided to replace its OPTOPUS (Lund, 1986) starplate system with
an automatic one, we opted for the arm approach. In that we were
following J. Hill and his MX ( Hill {\it et al.} 1986 ) and T. Ingerson and his
ARGUS (Ingerson, 1988), our neighbour instrument at CTIO. MX uses 
32 arms and ARGUS 24 arms. Our fibre positioner MEFOS, the Meudon ESO Fibre
Optical System (Gu\'erin {\it et al.} 1993), was built by the Observatoire de
Paris, France, under ESO contract.

\section{Overall Concept}
MEFOS is mounted at the prime focus, providing a large 
field (1 degree) for the multifibre
spectroscopy. The 3.6 m ESO telescope has a prime focus triplet
corrector delivering a field of one degree, the biggest for a
4m class telescope until the 2dF project will be in full operation 
at the AAT. The focal ratio is F/3.14, well suited for fibre
light input, leading to negligible focal ratio degradation (Gu\'erin {\it et al.} 1988). MEFOS 
(Gu\'erin {\it et al.} 1993) is put on the red triplet corrector 
with a trait-point-plan interface, allowing tilt correction, with 
respect to the optical axis. The telescope z movement of the prime 
focus is used for focusing. It
has 30 arms that can point within the 20 cm, one-degree field. 
One arm is  used for guiding, while the other 29 arms are dedicated
to the astronomical objects.                    
Figure 1 shows the general arms display. They are distributed around the
field as "fishermen-around-the-pond". The arms move radially and in
rotation: each arm can cover a 15 degree triangle
with its summit at the arm rotation axis and its base in the centre of the
field. Hence, each arm can access  objects at the centre of the 
field, whereas only one specific arm can reach an object at the field 
periphery. This situation 
changes  gradually from the centre to the edge of the field. Each arm has
its individual electronic slave board. The whole instrument is controlled
by a PC, independent of the Telescope Control System
(TCS). The arm tips -- shown on Figure 2 -- are carrying two fibres
separated by 1 arcmin. One is used for the object, the other for
the sky recording. Both are fed to the spectrograph. The
object and sky fibre, each 2.5 arcsec in diameter, can be exchanged and the 
fibre transmission cancelled.\\
\\ 

Coupled firmly to the arm tip is an imaging
fibre bundle that covers a 35 x 35 arcsec$^2$ area of the sky.
The image bundles are projected on a single Thomson 1024 x 1024 thick
Peltier cooled CCD, and are connected to the PC driving the arms.
The image bundle of the arms are moved to the object coordinates
and an image of the selected objects can be obtained.
Figure 3 shows the display of the galaxy field as seen through the 
29 windows of the CCD.\\ \\
\begin{figure}
{\psfig{figure=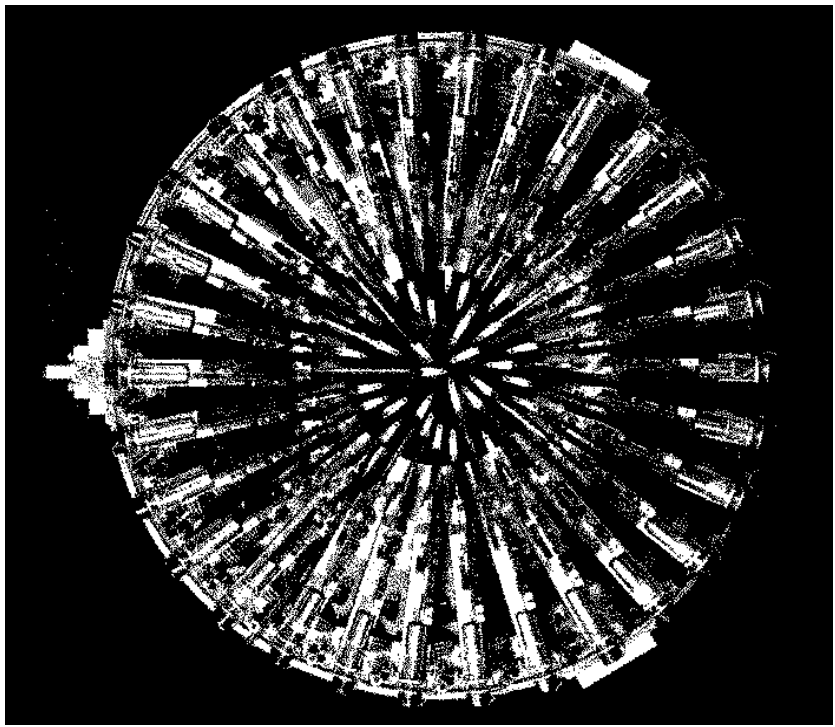,width=110mm}}
\caption{MEFOS general arm display around the one degree field 
and viewed from above.}
\end{figure}
\begin{figure}
{\psfig{figure=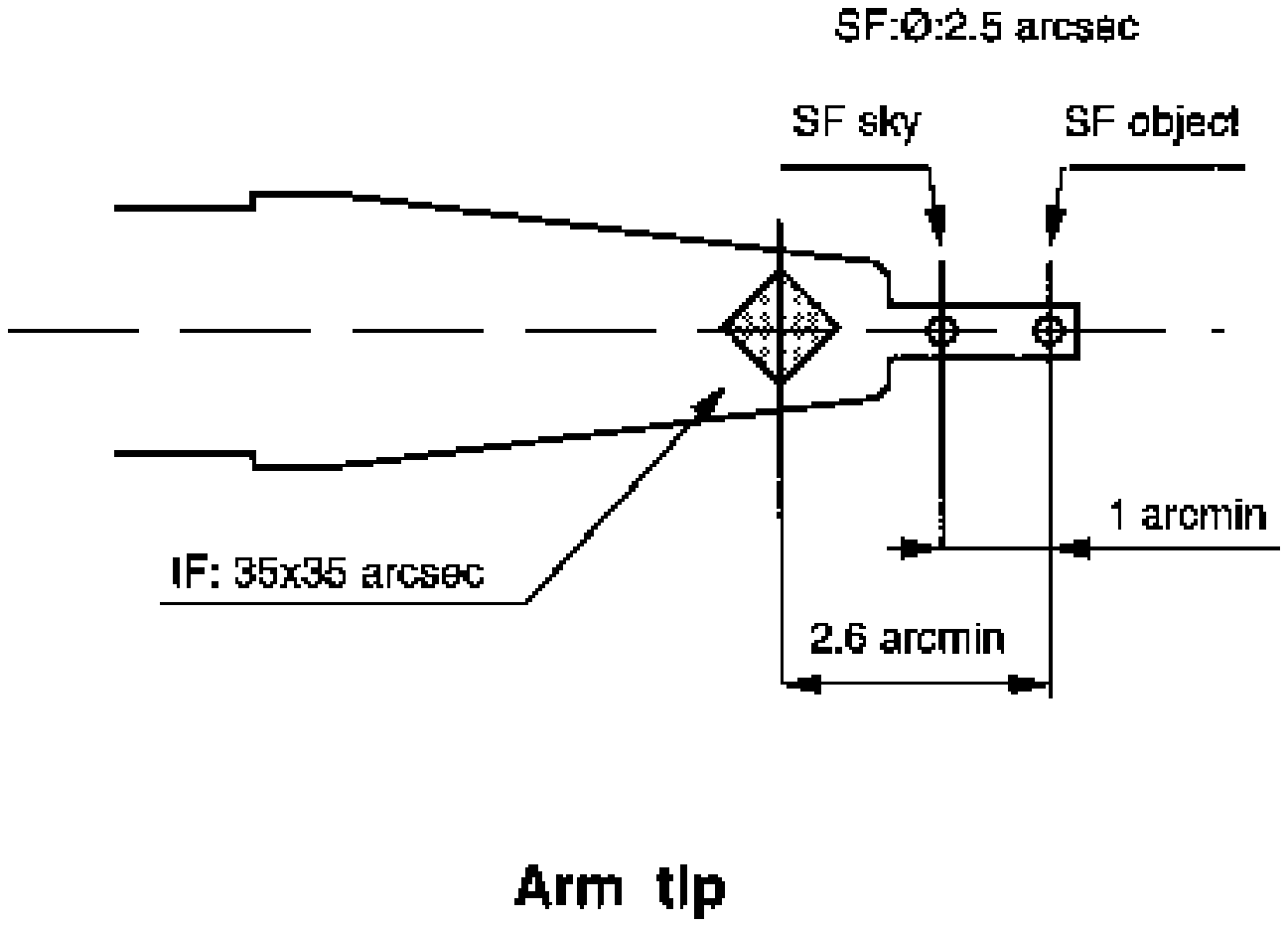,width=110mm}}
\caption{Arm tip showing the two spectroscopic fibres (SF) and the 
associated image fibre bundle (IF).}
\end{figure}
With this procedure, the objects 
on which the spectral fibre will be positioned can be seen in 
advance - contrarily to blind positioning. It is to our knowledge the only 
such instrument. By analysing the position of the object on the image,
in combination with the  fixed displacement between the image bundle 
fibre and the spectral fibre within an arm, a precise offset 
is determined and the arm sent to the optimal object position. This offset 
takes care of all imprecision due to the
telescope, the instrument and inaccurate coordinates. The poor
pointing of the telescope, and the fact that the corrector and the
instrument are frequently dismounted, would make blind positioning 
extremely dangerous. The final positioning accuracy, as measured on  stellar
sources, is 0.2 arcsec rms.

To avoid any motion between the telescope and the
instrument, the guiding is done in the arm environment. A special arm is 
dedicated for this task. This arm holds a large image conducting 
optical fibre with a field of view of 1.8 x 2.5 arcmin. To enlarge the 
field within which a guide star can be selected, this window can be 
positioned in 3 different (adjacent) positions.
The guiding camera is sensitive to V$<19$. Because of this faint 
magnitude limit, no guide star has to be defined prior to 
the observations. 
The guiding fibre can then remain 
in the outer position at the edge of the field, leaving more space 
to the positioning of the arms.

In the present stage, the spectral fibres, $135\micron$ 
in diameter and 21 m
long, are going from the prime focus down to the Cassegrain, where the B\&C
ESO spectrograph is located. This spectrograph is fitted with a F/3
collimator to match the fibre output beam aperture. It has a set of
reflection gratings with dispersion range from 2.3 nm/mm to 23.1 nm/mm, and a Tek 512 x 512 thin CCD. \\ \\

\begin{figure}
{\psfig{figure=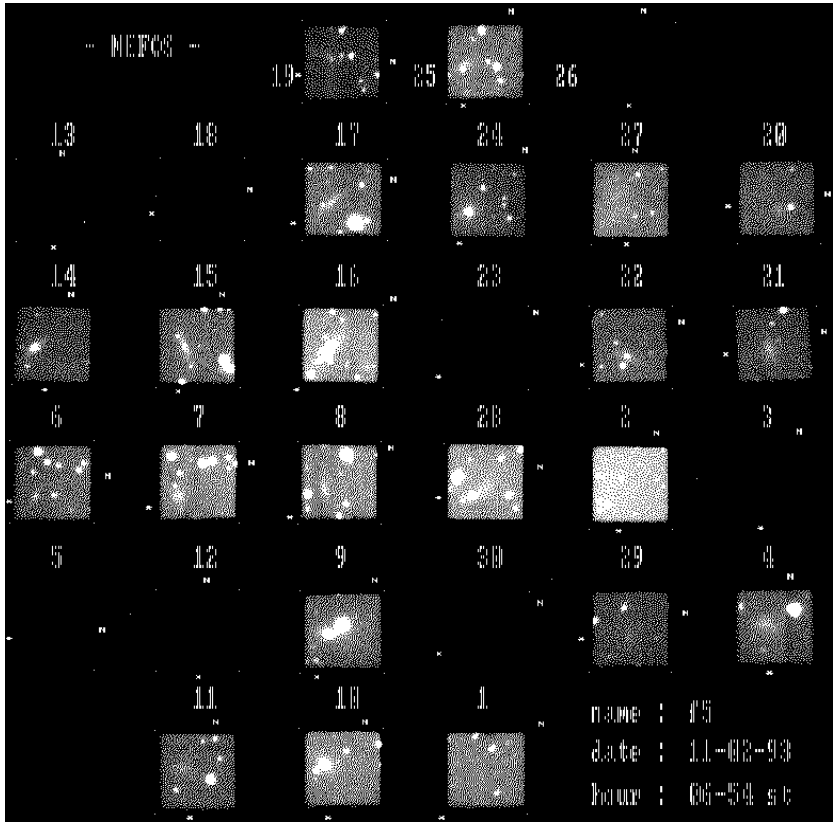,width=110mm}}
\caption{Hardcopy of the monitor screen, showing the sky field 
recorded by the Peltier cooled CCD of the 30 image fibre bundles after 5 
minutes integration. The stars and galaxies are clearly visible.}
\end{figure}

\section{Technical and Scientific Tests on Sky}

The first scientific observation was performed in February 1991
with a MEFOS prototype consisting of nine arms only 
(Bellenger {\it et al.} 1991). During this period a  
galaxy of magnitude B=$18\fm6$ was observed with a resolution of R=370.
The 1$^h$ exposure resulted in spectrum with a S/N=50 in the red part 
of the continuum (550 nm).

The first technical run in full configuration at the 3.6m telescope 
took place in October 1992. 
As only 2 1/2 of the 5 allocated nights were clear, a second run was
necessary for the testing and the commissioning. This was 
carried out in February 1993 during 5 clear nights. 
Some problems with the centering of the image fibres on the object 
coordinates were noted. It appeared that the optical aberrations
correction curve was not accurate enough. We had adopted this curve
as it was derived from photography at the prime focus.
The mounting of this instrument entails different
conditions. By moving an image fibre across the MEFOS field 
in real configuration, the correction curve could be adjusted.
The objects are now falling within a few arcseconds of the centre 
of the image fibre helping for their identification.

In the 7 nights following this run, MEFOS was used on 
two scientific programmes. The instrument worked well and little
time was lost due to technical reasons.
Cuby \& Mignoli (1994) studied sky subtraction procedures using
MEFOS with and without beam switching. They were able to measure 
redshifts for galaxies with magnitudes as faint 
as B=22$^m$ from a 90 min exposure (2 x 45 min). 
A B=$22\fm6$ galaxy was observed as well, but the resulting
signal to noise ratio of S/N =7 was too low for a reliable redshift 
measurement.

Since then, MEFOS was used in 84 nights, without a single
failure. Since April 1994, MEFOS has been used without any further 
support from the building team.

\section{Observing with MEFOS}

In order to compensate the building team of the DAEC for their efforts, 
32 nights of guaranteed time were allocated to scientists of this 
laboratory. From the submitted scientific proposals, four programmes were 
selected. Each was granted 8 observing nights with MEFOS.

Based on the experience gained in one of these programmes, we  
describe in the following the preparations before
an observing run as well as the typical procedures while
observing with MEFOS at the 3.6m telescope. 
In the next section we will then discuss a number of 
specific problems that we encountered while reducing the data,
and, at the end, a number of examples and some statistics will 
allow some insight into results that can be expected from 
multifibre spectroscopy with MEFOS.

The described procedures as well as the obtained data are, of course, 
strongly related to the selected project. It is therefore 
important to know some details about the project when discussing
the performance of MEFOS. The scientific project ``Search in the 
Galactic Plane toward the Great Attractor'' aims at mapping the 
galaxy distribution in velocity space in the Zone of Avoidance (ZOA) 
in the general direction of the Cosmic Microwave Background dipole (CMB) and the Great Attractor. 
Galaxies hidden behind the ZOA are important in derivations of 
the peculiar motion of the Local Group relative to the CMB 
, and in explaining the observed
streaming (bulk flow) motions, as e.g. 
in the Great Attractor region. 
Preliminary scientific results have 
already been presented (Cayatte {\it et al.} 1994, Kraan-Korteweg {\it et al.} 
1994, 1996a,b). This programme is a continuation of a long-term programme 
which made use of OPTOPUS, the previous multi-fibre spectrograph of ESO. 


\subsection{The Scientific Program}

A large part of the extragalactic sky is obscured by the Milky Way.
To bridge this gap, we have embarked 
on a search for galaxies behind the Milky Way to the low diameter 
limit of D=$0\farcm2$ and faint magnitude limit using existing sky 
surveys (Kraan-Korteweg \& Woudt 1994). 
To date over 10000 previously unknown galaxies were uncovered 
in the southern Milky Way ($265\deg \la \ell \la 335\deg, 
|b| \la 10\deg$). 

Three observational programmes are ongoing to determine the 3-dimensional 
distribution of the galaxies detected in the galaxy search:
multifibre-spectroscopy in the densest regions, long slit spectroscopy of 
more isolated galaxies, and HI-observations of fairly large, low 
surface brightness galaxies. The three approaches are complementary in
the depth of the volume they cover and the galaxy population
they are optimal for (cf. Kraan-Korteweg {\it et al.} 1994, 1995, 1996a).

After using OPTOPUS during 2 observing runs in 1990 and 1992, the 
multifibre spectroscopy was continued using the guaranteed scientific 
test time with MEFOS in 1993 (2 1/2 nights), 1994 (4 nights) and 
1995 (4 nights).

The magnitudes of the galaxies discovered in the optical search
range between $12\fm5 < B_J < 20\fm5$, with a strong peak 
$17\fm5 < B_J < 19\fm5$ (cf. Fig. 1 in Kraan-Korteweg \& Woudt, 1994). 
This range is well suited to the dynamic range of MEFOS.
Galaxies in this magnitude interval have an average
number density of the order of 100-200 galaxies$/\Box\deg$, suggesting a
more efficient use of OPTOPUS with 50 fibres in a half-degree field. 
However, the galaxies of this project lie behind the Milky Way. They 
are heavily obscured: around one magnitude at the
border $|b| = \pm10\deg$ of our survey, increasing to 5-6 magnitudes
towards lower latitudes (close to the Galactic dust equator 
the galaxies are fully obscured). The average number density
for $18^{th}$ magnitude galaxies is therefore considerably lower. In fact, 
the multifibre capacity of MEFOS with 29 objects per one-degree field 
in combination with its capability of obtaining good S/N spectra 
for these faint obscured galaxies of our survey, make
it the ideal instrument to trace large-scale structures close to 
the plane of the Milky Way.

The galaxies of the Zone of Avoidance survey not only are very obscured
and often of low surface brightness, but since they are
located in the Milky Way, the fields are crowded with foreground
stars, many of which are often superimposed on the faint galaxy images. Here, 
the imaging facility of MEFOS for object identification and optimal 
centering of the spectral fibre is again a most advantegous property (as 
for spectroscopy of LSB objects and/or in crowded fields in general). 

\subsection{Observational Set-Up}

\subsubsection {Arms to objects allocation}
To prepare an observation, it is necessary to have a list of objects with
coordinates precessed to the date of observation and known with a
precision of the order of one arcsec. Arm-object assignation software
is provided which can be obtained by ftp from ESO. This work is optimally 
done at the home institute before the observing run and does not cost 
telescope time. The software is also implemented at La Silla, allowing
new or optimization of field assignations during the observing run.

The arm assignation software displays the objects of an earlier 
prepared list on the screen. 
A circle of one degree (the size of the MEFOS field) can be
moved around the screen until optimal centering of the field is 
attained. The computer then suggests a possible configuration for an
arm-object assignation.  The software does take
the constraints of size and moving capacity of the arms into account, 
to avoid collisions. The smallest allowed distance between objects 
depends on the relative arm positions. It lies between 2 and 
5.8 arcmin. If the proposed assignation is not optimal from a scientific
point of view, it can be modified interactively (within the above
named limits).
Figure 4 shows an example of an assigned field. The arms are
displayed with their true extent.\\ \\
\begin{figure}
{\psfig{figure=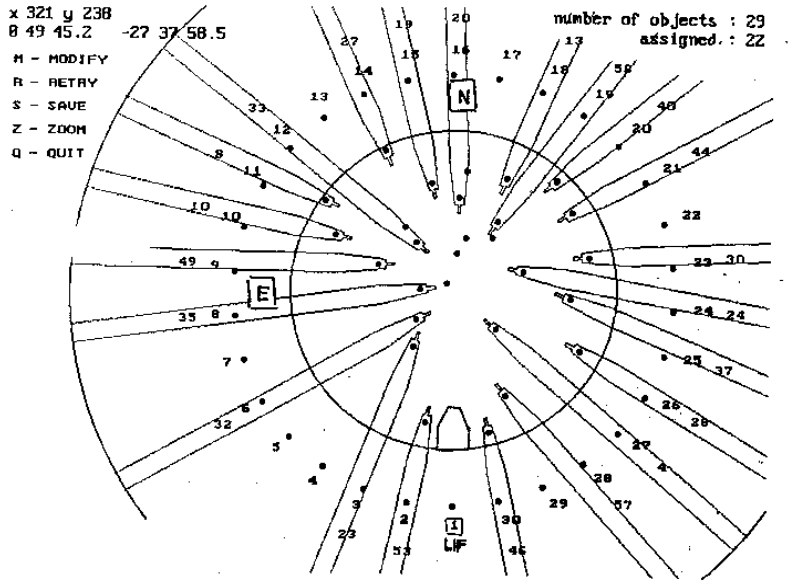,width=110mm}}
\caption{Example of an arm to objects field allocation. The inner 
circle limits the one degree field. The inner numbers are arm numbers, the 
outer ones are the object numbers given in the prepared list.}
\end{figure}

\subsubsection{Instrument set-up at the telescope}

MEFOS is installed with the corrector at the prime focus.
The North-South direction has to be
known with a high precision. A set of star fields, equally
spaced in right ascension, are used during the set-up for calibration. 

The pointing of the telescope is essential when observing with
MEFOS. During the set-up, the pointing will be checked in detail. This
will lead to a pointing model which is automatically applied during the
whole observing run. It is therefore not necessary to recheck the
pointing. Any change in the pointing should be avoided, as this will
destroy the pointing model. The pointing only has to be rechecked if 
the system had, for some reason, to be rebooted.

The instrument is focussed by pointing the guiding arm 
to a star. This fibre has a larger field of view compared to the object 
image fibres. 

\subsection{Observational Procedure}

\subsubsection{Objects acquisition}

The telescope is pointed towards the selected field. As soon as the
the telescope is locked on the centre of the MEFOS field, the guiding 
has to be switched on. 

There are 3 adjacent guide fields of 1.8x2.3 arcmin each, 
in which a  guide star can be found. The magnitude of the star can be 
as faint as 19th magnitude. 
In the exceptional case that no guide star can be found
in any of the 3 possible guide field positions, it is better to switch 
to another target field in order not to lose valuable observing
time. An arm-object reassignment to enhance the likelyhood of finding
a star in the guiding window can be done at a later point of time.
The arms are then sent to their previously assigned object positions, 
with the image fibre bundles centered on the objects.
During the arm positioning, the PC controlling the MEFOS operations,
continously displays the movements of the arms. It typically takes 
about 4 minutes for the arms to reach their final positions. 
As soon as the arms are in position, a CCD image is taken 
(see Fig. 3). The images are crucial, because they 
allow object identification and accurate centering of the spectral fibre. 
These images are raw and cannot be used for other scientific purposes 
like photometry. The readout time for the CCD image is 17 sec. The 
noise is quickly sky dominated. An exposure time for a V=17$\fm$2 star
is of the order of 20 sec with  1.7 arcsec seeing. For a B=20$^m$ galaxy, 
depending on the shape, about 3 min are needed. 

We generally exposed as long as 5 or 7 minutes because the objects of 
our programme are obscured galaxies, often of low surface brightness in 
fields crowded by foreground stars. 
The identification of our objects in the 35"x 35" field was therefore often
difficult. For brighter objects in less crowded fields the advocated 
2 to 3 minutes is adequate.
 
\subsubsection{Object identification}

After the images are taken and displayed on the screen, the observer 
has to identify the object on the respective images for final optimal
positioning. This can be done automatically or manually. An internal
routine searches for the brightest objects within the images and 
puts numbered boxes around them. The correct object can then be 
identified by entering the number of the box into the computer. 
The optimal position can also be indicated manually by clicking 
with the cursor on the required position. The latter is a
more efficient procedure for centering of very low surface brightness
objects. Depending on the number of objects in the field of view and on 
the surface brightness of the objects, this step takes between 5 and 10 
minutes. If the objects are very faint or multiple, it is advisable to 
make polaroid copies and mark the position where the spectral fibre 
will be put. It takes about one minute to send the spectroscopic fibres 
to the recentered objects.

\subsubsection{Calibration and scientific exposures}

After the identification step, the observing procedure depends on the 
scientific programme. A flat field with 
an internal white lamp allows to identify and to check the 
position of the fibres on the CCD. The exposure time is about 10 sec.
The He and Ne lamps are used for the wavelength calibration 
(note that the Neon lamp has to be requested before the setting). 
It is recommended to have a filter (BG 28) put in front of the Neon 
to reduce the intensity of the red Neon lines and avoid saturation. 
The two lamps can be exposed simultaneously. Based on our experience,
an exposure time of about 1 min is reasonable.                          

Then the scientific exposures can begin. Depending on the programme,
magnitude and surface brightness of the objects,
a sequence of 30 min exposures follows.
With more than two consecutive exposures of one field, another 
calibration was taken to test the stability of the wavelength 
calibration. 

This whole procedure has to be repeated for every field. The overhead
between one scientific exposure of a field to the next field
typically takes 20 to 30 min, calibration included.
With two observers at the telescope some of these operations 
can be done simultaneously, saving therewith on 'real' observing time.

For an exposure time of 2x30 min, 5 to 6 fields per night 
can be observed depending on the length of the nights.
We observed 9 fields in  2 1/2 nights in February 1993, 19 
fields in 4 nights in February 1994 and 19 fields in 4 nights in May 1995. 

\section{Data Reduction}

We used the CCD Tek \#32 in our observing runs with the grating
\#15. This resulted in a dispersion of 170\AA/mm and a resolution of
about 11\AA. The wavelength range was chosen to be 3850\AA - 6150\AA.

The data reduction has been carried out on Sun stations with
the IRAF package.
The reduction process consists of 
subtraction of a mean bias, extraction of the spectra, 
wavelength calibration, fibre transmission correction, 
sky subtraction, removal of cosmic-ray events and 
redshift determination. For the last step we used the RVSAO package 
obtained from the Smithsonian Astrophysical Observatory  
developed by Doug Mink. With this package,
radial velocities can be obtained with the cross-correlation method 
proposed by Tonry and Davis (1979), and redshifts can also be determined 
from emission lines if present. In the course of our observing runs, we 
regularly observed a number of standard velocity stars (1-2 per night)
which were subsequently used in the cross-correlation of each object
spectrum. 

We will not give a detailed description of the various steps
of the the reduction process -- this will be given in the forthcoming
paper on the scientific programme -- but rather highlight  
some specific problems. 

\subsection
{The determination of the spectra position and the wavelength calibration}

Efficient extraction of spectra and accurate wavelength 
calibration are critical in measuring accurate redshifts. This
does require deep understanding of the performance of 
the system such as the illumination of calibration 
and flat-field sources. In the following we point out some problems 
that we encountered while reducing the multifibre data.

\subsubsection{Extraction and calibration procedure}

For each fibre the position of the spectrum on the CCD has to be defined.
This is done by measuring the centroid of the profile perpendicular
to the dispersion. The curvature in the dispersion
direction is corrected by fitting the centroid position for each
object as a function of the coordinate along the wavelength axis.
The positions of the spectra are determined from a flat field
exposure in the same telescope configuration
as the objects of the field exposure.
No "cross-talk" between two adjacent spectra has been evidenced, so an
extraction window as large as the spectra separation can be chosen. In
our case we use an extraction window of 5 pixels for each aperture
(adding faint signal in the profile wings reduces the S/N ratio for the
object spectrum). Spectra for each aperture in each exposure were then
extracted with "variance" weighting.

The dispersion was determined for 
each aperture (e.g. each spectrum) from the corresponding aperture on 
the He-Ne lamp calibration exposures. 
For the dispersion solution we used 13 emission lines from the He-Ne lamp.
We fitted a cubic spline of order 3 to the non-linear component. The
residuals were not larger than 0.3\AA\ . By measuring the scatter in the
skyline position, we always obtained an uncertainty in the wavelength 
calibration below 0.5\AA\ (e.g. 30$\kms$).

\subsubsection{Flexure and stability}

The total flexure between the CCD mounted on the B\&C spectrograph
(mounted at the Cassegrain focus) and the fiber focus has been checked in both
directions. To look for positioning repeatability of 
the spectra, we have compared two flat field exposures of different 
telescope pointing. After both the telescope and all arms were 
moved to other locations, differences between the positions 
of the spectra for two flat-field exposures were found,
though never more than 0.25 pixels (e.g. 7\micron). 
Flexure with telescope tracking was checked by comparing two 
flat field exposures taken with the same telescope pointing.  
The maximum noted flexure in an hour never exceeded 0.25 pixels, i.e. 7\micron 
.

In the dispersion direction we compared two lamp calibration 
exposures taken before and after successive field exposures. 
The 1993 MEFOS run revealed a shift of about 1\AA\ per hour 
depending on telescope position. This corresponds to a quarter 
of a pixel, i.e. is of similar order as in the spatial direction.
This effect was seen in almost all the fields. It could 
be due to a continuous drift with time. 
The February 1994 MEFOS run did not exhibit any such effect.

\subsubsection{Positioning and calibrating problems}

While reducing the fields of the February 1993 and 1994 runs, we 
noticed inconsistencies between the positions of the spectra obtained
from the flat field as compared to the objects:
the difference in position between the flat and the object 
is roughly constant up to the middle of the CCD frame and then 
decreases as a function of pixel position in the spatial direction up 
to -0.5 pixels (e.g. 13\micron). This trend is displayed in
Figure 5. Although the mean curves of the different fields
are slightly displaced from each other, they all clearly  
exhibit the same trend. Moreover the mean shift changes as a
function of wavelength. It is stronger in the blue part of the
image. The problem therefore does not seem related to the mechanical 
stability and reliability, but rather to the difference in the light 
path of the lamp used for the flat field exposure and the object exposure.
Cuby and Mignoli (1994) noticed displacement problems as well.
However, they find a difference between flat-field and sky 
spectra positions which varies linearly with the centroid of the
spectra.
\begin{figure}
{\psfig{figure=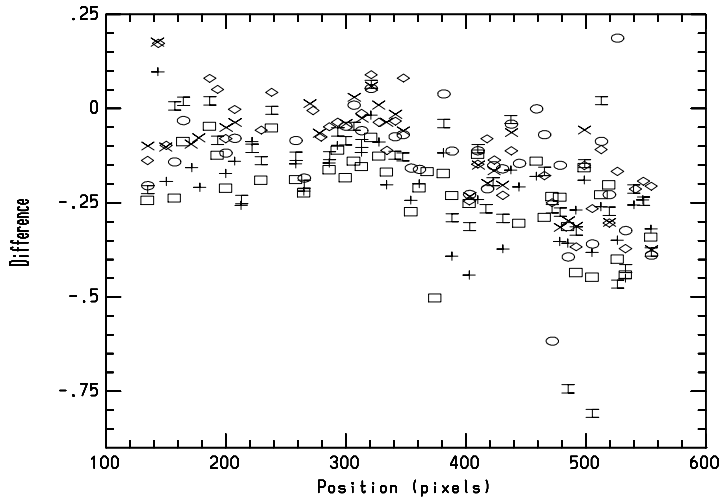,width=110mm}}
\caption{Difference in the position between the flat and the object
as a function of pixel position in the spatial direction on the CCD. The different
symbols represent six of the fields from the 1993 run.}
\end{figure}

Because of these difficulties, the MEFOS-user is adviced 
to check for such systematic effects and not 
to blindly use the flat fields for defining the spectra positions.
To circumvent these problems, we extracted the spectra positions directly 
from the object spectra. The two exposures per field
were summed up if no shift between the positions on 
the CCD frames could be found. This addition increases  
the signal to noise ratio in the less sensitive, blue part of the spectra.
The errors in the position of the spectra in the faint blue part
are of the order of 0.3 pixels (as compared to the 
FWHM of the spectra profile of about 3 pixels). 
A large extracting window, on the same order of the spectra 
separation, will decrease the loss of signal due to the imprecise 
determination of centroid of the profile. \\

For certain fibres 
the spectra were not properly wavelength calibrated. For these 
spectra the [OI] skyline wavelength position was not close to the 
actual value of 5577.4\AA\ .
So we also checked the alignment of the fibres along the slit by 
comparing He-Ne lines positions for different apertures; 
the result is shown in the bottom panel of Fig. 6 : the mean 
spectral-shift of all the He-Ne lines in the dispersion direction has 
been plotted versus the aperture position coordinate on the spatial axis. 
The points seem to be located around a parabola which 
represents the image of the slit through the spectrograph. 
In the top panel of Fig. 6 , we have plotted the differences between the measured position of the [OI] skyline and 5577.4\AA\ versus the aperture position coordinate on the spatial axis. 
The largest displacement of the skyline position occurs for the 
fibres which are the most misaligned on the slit. As suggested 
in the previous paragraph, different optical paths for sky and 
lamp exposure are suspected. Here we observe this effect in the dispersion 
direction. We cannot wavelength-calibrate the spectra correctly for
objects observed with fibres that are located far from the slit center in 
the spectral direction. \\
\begin{figure}
{\psfig{figure=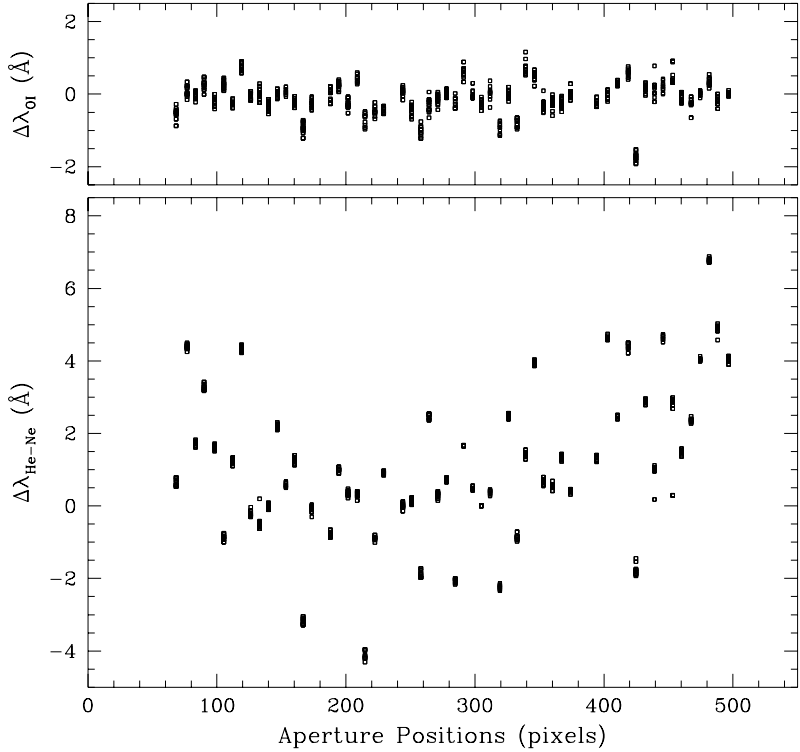,height=100mm}}
\caption{Mean shift of all the He-Ne lines in the dispersion 
direction as a function of the aperture position in pixels on the other axis 
(bottom panel). The difference between the measured and the real wavewlength of the [OI] skyline is presented in the
top panel.}
\end{figure}
The optical paths difference could be explained if 
misaligned fibres have undergone some bending effect or that the
plane of the slit (in front of the fibers) is not perfectly perpendicular. This would
change the position of the image if the CCD is illuminated 
by the sky or by a calibration lamp.

To correct the bad wavelength calibration we estimated a mean 
shift by averaging the [OI] skyline offsets. These were measured 
from each exposure in a field. We then applied an inverse shift 
of the corresponding amount to each spectrum 
taken in the "misaligned" fibres. The largest discrepancies between measured and actual value was consistently seen for the same fibers in all the 18 observed fields of the 1994 run. The r.m.s error of the mean 
skyline offset was never larger than 0.17\AA\ . Generally, it was
of the order of 0.12\AA\ , corresponding for a wavelength of 5577.4\AA\ 
to an r.m.s. error of 6$\kms$ in velocity.

\subsection{The fibre transmission correction and sky subtraction}

We estimated the 
relative transmission of each fibre from the [OI] 5577.4\AA\ skyline flux to correct the fiber efficiencies.
After this correction, we averaged all the sky spectra 
in a given field. This minimizes the noise of  
the sky spectrum to be subtracted from the object spectra.
With MEFOS, each arm carries its own sky fibre, allowing therewith
other sky subtraction methods. Detailed investigations on optimal 
sky subtraction routines have been performed by Cuby and Mignoli (1994). 
They find averaging over all sky spectra, after fibre efficiency 
correction as determined from the skyline flux the best solution.
This does of course depend on the programme. For spectrophotometry 
of galactic or extragalactic objects, for instance, precise 
flux calibration will be required.

The fibre relative transmission coefficient deduced from the [OI] 5577.4\AA\ 
skyline flux takes into account the sky variations over the field 
together with the effective fibre efficiency determination, without the 
possibility of disentangling the two effects. Generally, sky variation is 
rather weak below 6000\AA\, even on a one degree field (see results of 
Cuby and Mignoli, 1994). The transmission variations observed 
for some fibres during the night as well as from night to night, could 
therefore be largely due to a change in fibre efficiency, which can, for 
instance occur when light injection conditions vary. Nevertheless, the 
transmission coefficient values obtained from two exposures on the same 
field show very little scatter. From the statistic on the fibre transmission 
difference between the two exposures we derived an r.m.s. error on the 
estimation of this coefficient below 2\%.

\section{Results}

\subsection{Objects per field}
In 1993 and 1994, 506 objects have been observed on 27 fields. This 
translates to an average of 19 objects per 
MEFOS field. This number is considerably lower than the maximum 
of 29 objects within a one degree circle. 
On the one hand, this reflects the choice of a field, 
on the other hand the constraints of the instrument itself. 
Besides scientific priorities, constraints are given by 
the arms; i.e. the width of the arm  and the minimum allowed distance. 
The average number of objects that we could observe in crowded fields 
was of the order of 22 to 24 galaxies. Numbers of 27 to 29 were indeed 
achieved if the galaxy density reached a hundred or more, as in 
fields of the cluster Abell 3627 (Kraan-Korteweg {\it et al.}, 1996a).  

\subsection{Derived redshifts}
Of 473 spectra in 24 reduced fields, redshifts have been obtained for 93\% 
of the objects, among which 84\% are of galaxies and 9\% are of stars.\\
The perturbations of the galaxy spectrum by foreground stars (of the 47 spectra 
contaminated by a superimposed star, only 6 redshifts of galaxies could be 
obtained) is, of course, a problem inherent to our programme and should not 
affect observing programmes away from the Galactic Plane. 
Among the redshifts obtained for galaxies, 77\% are reliable (i.e. with 
a S/N ratio larger than 20 at 5000\AA\ ) 
with errors in velocities which in the mean 
are about $60 \kms$. For 22\% of the objects, the redshift could be 
attained from emission lines.  

The distribution of the good S/N redshifts are displayed in Fig.
7 for the velocity range of $0 < v_h < 45000\kms$. 

The distribution in Fig. 7 depends strongly on the extragalactic large-scale 
structures that are being intercepted in the Milky Way. For instance, 
the strong peak at about $5000 \kms$ mainly stems from 4 MEFOS fields 
centered on the cluster A3627 at $(\ell,b)\approx (325\deg,-7\deg)$. 
The velocities from these 4 fields are marked by the dashed histogram. 
In fact, the velocities from these observations, in combination with the
result from our complementary observations at the SAAO and at Parkes,
have proven this cluster to be very massive and therefore to be at the 
bottom of the potential well of the Great Attractor 
(Kraan-Korteweg {\it et al.} 1996a).\\
\begin{figure}
{\psfig{figure=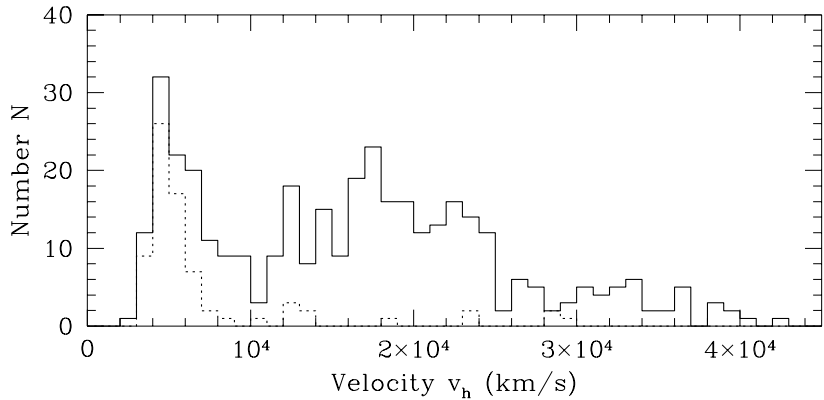,width=110mm}}
\caption{Distribution of the 334 redshifts obtained from observations
with MEFOS in 1993 and 1994 from good S/N spectra.
The dashed histogram is a subsample of 4 fields centered on A3627,
a massive cluster close to the core of the Great Attractor.}
\end{figure}
The mean velocity of the observed sample is $<v_h>=16500\kms$. 
The peaks at 12000, 18000 and 23000$\kms$ are part of filaments
that diagonally cross the Zone of Avoidance. They can be traced 
to higher latitudes on both sides of the Milky Way and are 
indicative of extragalactic large-scale structures of
considerable size (a few hundred Mpc). For further details see
Kraan-Korteweg {\it et al.} (1994 \& 1996b). \\
\begin{figure}
{\psfig{figure=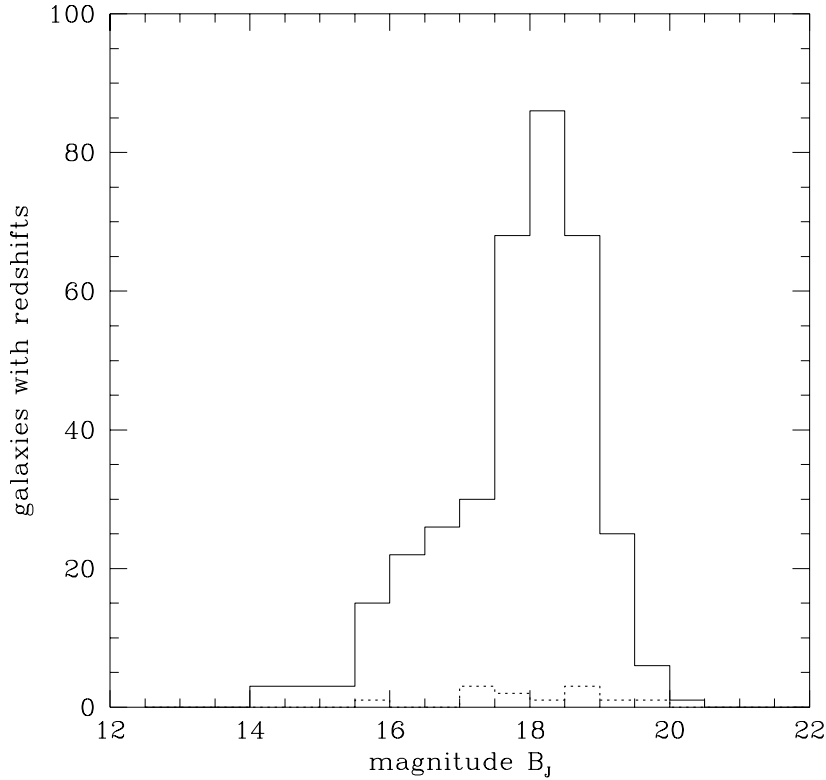,width=110mm}}
\caption{Magnitude distribution of the galaxies for
which redshifts have been determined from a 2x30 min exposure
with MEFOS. The histogram reflects the magnitude distribution
of the galaxy sample rather than a constraint of the instrument
MEFOS. This can be seen from the interior dashed histogram
which are galaxies for which no redshift could be determined
because of low S/N ratio.}
\end{figure}
Here, it suffices to 
maintain that MEFOS produces good quality
spectra for galaxies as distant as $45000\kms$ . 

\subsection{Dependence on magnitude and mean surface brightness}
The above results do depend on the properties of the observed
galaxies like the magnitude and the mean surface brightness, 
although the best parameter would be the central surface 
brightness corresponding to the flux collected at the position of 
the fiber. One can imagine a very low mean surface brightness galaxy 
with a relatively bright nucleus, giving a good S/N spectrum. 

In Fig. 8 the distribution of the apparent magnitudes B$_J$ of 361 
galaxies with redshifts is displayed. 
The magnitudes are estimates from the IIIaJ film copies of
the SRC. Although eye estimates, it has been shown that these 
magnitudes are accurate to $\sigma=0\fm5$ with no deviations from 
linearity up to the faintest galaxies. (cf. Kraan-Korteweg and Woudt, 1994, 
Kraan-Korteweg 1996).  

The mean magnitude of the observed galaxy sample is $<B_J>=17\fm8$.
If this distribution is compared to the one of the galaxies for which 
no redshift could be extracted (dashed histogram in Fig. 8), no clear 
bias is visible toward the faintest objects. Therefore the derivation 
of a redshift in this magnitude range does not critically depend 
on magnitude. The distribution shows a strong peak between $17\fm5$ 
and $19\fm0$ which does reflect the magnitude distribution of the Galactic
Plane survey (cf. Fig. 1 in Kraan-Korteweg \& Woudt, 1994).
\vspace{0.4cm}\\
A slightly stronger dependence has been noted with mean surface brightness
$SB_J$. The mean surface brightness ranges between $22\fm8 \le SB_J \le 24\fm5$
with a narrow Gaussian distribution around $<SB_J>=23\fm5$ ($\sigma=0\fm29$), 
whereas the galaxies for which no redshift could be 
determined have a mean of $<SB_J>=23\fm8$ ($23\fm3 \le SB_J \le 24\fm1$). 
But a redshift could be derived for the faintest few galaxies as well 
as for the one with the lowest surface brightness. 

A weak correlation between velocity and surface brightness was noted:
while the galaxies with $v<30000 km/s$ show a stable distribution about 
the mean, the galaxies with $v>30000\kms$ reveal a decrease in mean
surface brightness with increasing redshift.
\vspace{0.4cm}\\
In Figure 9, a few characteristic examples of galaxies from our 
survey and their resulting spectrum from a 2x30 min exposure with MEFOS 
are shown. They illustrate the quality of the obtained 
spectra as a function of the properties of the observed galaxies and 
visualize the above discussed tendencies. 
Three of the four objects presented in Fig. 9 have a $B_J$ 
magnitude nearly equal to the mean one of the whole sample, but the
mean surface brightness spans a large range. The object shown in the lower-left panel 
is one of the faintest galaxy of our sample, a 20$^{th}$-magnitude 
spiral galaxy with a very low mean surface brightness ($SB_J=24\fm1$). 
The CCD images of the galaxies clearly show that the central 
surface brightness is the most important factor in achieving a high 
S/N ratio. The velocities of the two objects in the upper panels 
are of the order of the mean velocity obtained for the sample,
but the galaxies in the lower panels are more distant. \\
\begin{figure}
{\psfig{figure=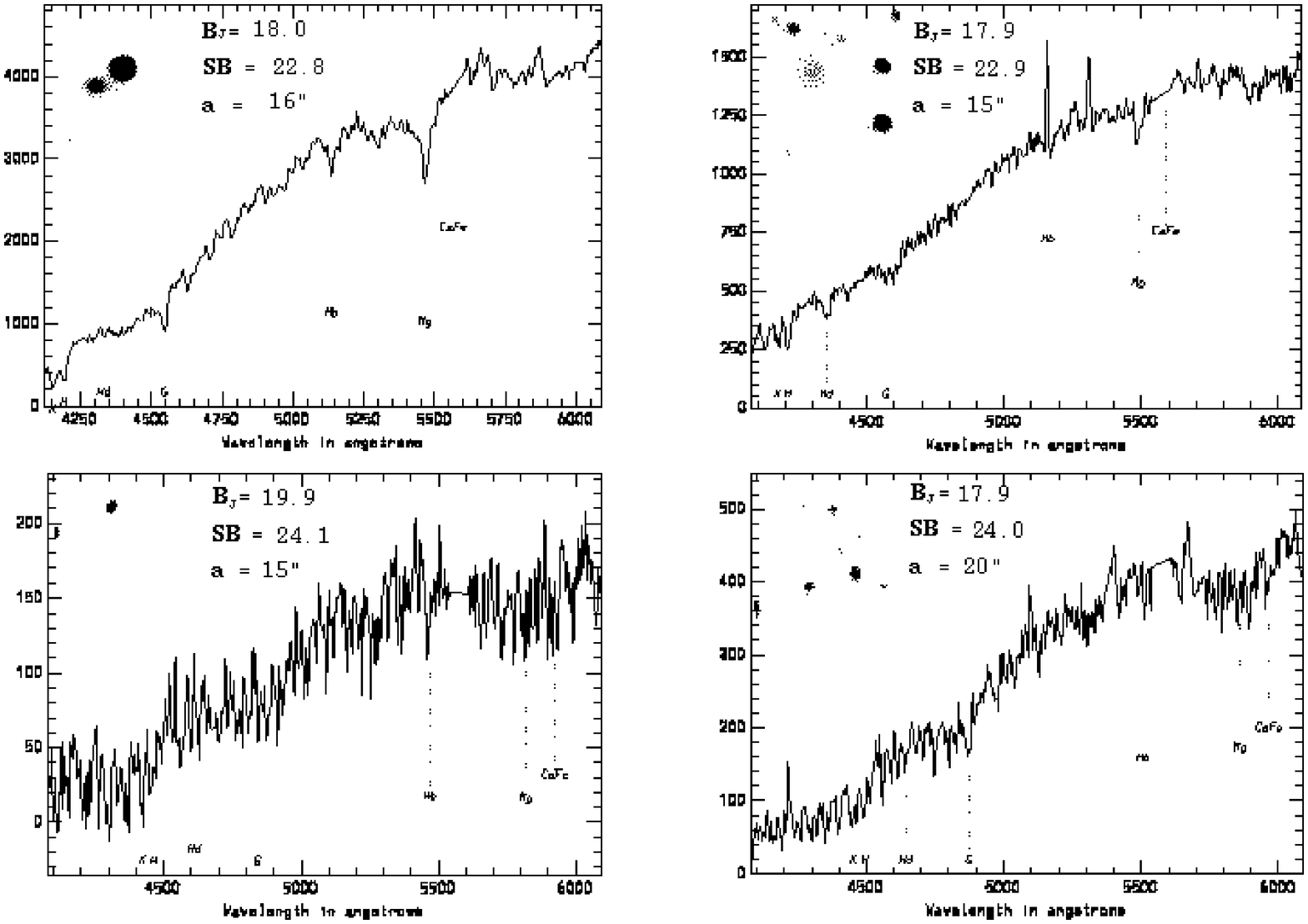,width=110mm}}
\caption{Spectra of four galaxies observed with MEFOS. The 0'.5x0'.5 CCD
images is shown above each spectra. Their magnitude B$_J$, mean
surface brightness and diameter are indicated.}
\end{figure}
>From the overall analysis of the observed sample, it can be maintained 
that good signal to noise spectra can be obtained with MEFOS from
2x30 min exposures for galaxies up to a magnitude of 
$B_J\le19\fm5-20\fm0$, the approximate magnitude limit of our survey.
For these obscured galaxies behind the Milky Way,
we have measured redshifts of galaxies up to $v_h< 45000\kms$. 
Reliable redshifts can be expected for a percentage of over $\ga 90\%$. 
This efficiency does depend on surface brightness and, especially in the 
case of this programme, is limited  by interfering foreground stars, an effect 
which would be much stronger were it not for the imaging capability of MEFOS. 
\\

\section{Conclusions}
In this paper we present the features of MEFOS, a multi-object fibre 
positioner coupled to the B\&C spectrograph and implemented at the prime 
focus of the ESO 3.6 m Telescope at La Silla. This instrument is open 
to the ESO community and was used on the sky for 84 nights, until 
now. MEFOS is an arm type instrument which is able to position 29 fibres 
on designed targets and has the same number of sky fibres. Very precise 
fibres positioning with this instrument is only possible on this generation 
of telescopes with a vision system that is associated to the arms. 
MEFOS aim is the observation of moderately crowded fields that need a one 
degree field for efficient observations.

Several scientific programmes have been conducted with MEFOS. The results 
obtained for one of them, aiming at measuring redshifts of galaxies found 
in the Galactic Plane, is given as an example of what can be observed 
with MEFOS, and what were the encountered problems with the data 
reduction.

\acknowledgements
{We are indebted to M. Dreux who built, with the SYNAPS company, the MEFOS
CCD camera and made it run on the telescope. R. Bellenger and R. Schmidt
designed and built the arm electronics, with the RAISONANCE Company, which
worked with no failure during all the telescope runs. The mechanical
design, machining and assembly was made by  the Service des Prototypes of
CNRS at Bellevue. Without all those people skill and total involvement, 
MEFOS would never been born. 
We also wish to thank Patrick Woudt for his participation to the reduction 
of the 1994 run data and for providing the Figure 6.}

\end{document}